\documentclass[manuscript]{aastex}

\def\lax{{$\mathrel{\hbox{\rlap{\hbox{\lower4pt\hbox{$\sim$}}}\hbox{$<$}}}$}}
\def\gax{{$\mathrel{\hbox{\rlap{\hbox{\lower4pt\hbox{$\sim$}}}\hbox{$>$}}}$}}

\usepackage{color}

\def\vec#1{\mbox{\boldmath $#1$}}
\usepackage{lscape}
\slugcomment{}

\begin{document}
%% LaTeX will automatically break titles if they run longer than
%% one line. However, you may use \\ to force a line break if
%% you desire.
\title{Radiation-Driven Warping of Circumbinary Disks Around 
Eccentric Young Star Binaries}
%Eccentric Binary Young Stars}
%Eccentric Binaries in Young Star Forming Regions}

 \author{
 Kimitake \textsc{Hayasaki}\altaffilmark{1},
 Bong Won \textsc{Sohn}\altaffilmark{1,2},
Atsuo T. \textsc{Okazaki}\altaffilmark{3},
Taehyun \textsc{Jung}\altaffilmark{1},
Guangyao \textsc{Zhao}\altaffilmark{1},
   and
 Tsuguya \textsc{Naito}\altaffilmark{4}
   }
\altaffiltext{1}{Korea Astronomy and Space Science Institute, Daedeokdaero 776, Yuseong, Daejeon 305-348, Korea}
\email{kimi@kasi.re.kr}
\altaffiltext{2}{Department of Astronomy and Space Science, University of Science and Technology, 217 Gajeong-ro,
Daejeon, Korea}
\altaffiltext{3}{Faculty of Engineering, Hokkai-Gakuen University, Toyohira-ku, Sapporo 062-8605, Japan}
\altaffiltext{4}{Faculty of Management Information, Yamanashi Gakuin University, Kofu, Yamanashi 400-8575, Japan}
%%%
%
\begin{abstract}
We study a warping instability of a geometrically thin, non-self-gravitating, 
circumbinary disk around young binary stars on an eccentric orbit. Such a disk 
is subject to both the tidal torques due to a time-dependent binary potential and 
the radiative torques due to radiation emitted from each star. 
The tilt angle between the circumbinary disk plane and the binary orbital plane 
is assumed to be very small.
We find that there is a radius within/beyond which the circumbinary disk is unstable to 
radiation-driven warping, depending on the disk density and temperature 
gradient indices. 
This marginally stable warping radius is very sensitive to viscosity parameters, a fiducial disk 
radius and the temperature measured there, the stellar luminosity,  and the disk surface density 
at a radius where the disk changes from the optically thick to thin for the irradiation from the central 
stars. On the other hand, it is insensitive to the orbital eccentricity and binary irradiation parameter, 
which is a function of the binary mass ratio and luminosity of each star. 
Since the tidal torques can suppress the warping in the inner part of the circumbinary disk, 
the disk starts to be warped in the outer part.
While the circumbinary disks are most likely to be subject to the radiation-driven warping on a AU 
to kilo-AU scale for binaries with young massive stars more luminous than $10^4L_\odot$,
 the radiation driven warping does not work for those around young binaries with the luminosity 
 comparable to the solar luminosity.
\end{abstract}
\keywords{accretion, accretion disks - hydrodynamics - masers - protoplanetary disks 
- stars: massive - stars: low-mass - stars: formation - (stars:) binaries: general} 
%
%%%%%%%%%%%
\section{Introduction}
%%%%%%%%%%%
%
% 1st paragraph
%
About 60$\%$ of main sequence stars are considered to be born as binary or multiple 
systems \citep{dm91}. Indeed, the direct imaging of the circumbinary disk was successfully 
done for a few young binary star systems with the binary mass comparable to a solar mass, 
including GG Tau \citep{dut94} and UY Aurigae \citep{duv98} by the Plateau de Bure 
Interferometer (PdBI). It is confirmed by numerical simulations that the circumbinary disk is 
formed around such young low mass binary stars embedded in the dense molecular gas 
\citep{al94, bb97, gk1, gk2}.

%
% 2nd paragraph
%
The existence of a significant number of OB eclipsing binaries \citep{hi05} suggests that 
most known massive stars are also born as binary or multiple systems \citep{sana12}. 
Such binaries are expected to be accompanied by circumbinary disks at the early stage 
of the massive star formation. Contrary to the case of T Tauri stars, however, the massive 
star formation is still poorly understood because the observation of massive star forming 
regions is challenging. This is because they are obscured by a dusty environment and the 
chance to observe massive young stellar objects is small because of their short lives.
\cite{araya10} suggested that observed periodic maser emissions are originated from 
periodic accretion onto the binary stars on a significantly eccentric orbit from the circumbinary 
disk \citep{al96}, although there are two other scenarios: the colliding wind binary scenario 
\citep{vdw11} and the scenario of pulsation of protostars growing via high mass accretion 
rate \citep{ina13}.

%
% 3rd paragraph
%
It is natural that the circumstellar and circumbinary disks in the star forming regions have 
a warped structure. \cite{kr05} suggested that there is observational evidence for the disk 
warping in GG Tau system. The tidal effects of close encounters between stars and a disk 
surrounding it can make the disk outer parts misaligned or warped \citep{mb06}. If the 
circumbinary disk is originally misaligned with the binary orbital plane, it should be warped 
by the tidal alignment, although the disk tilt secularly decays by the disk viscosity \citep{flp13,lf13}. 
The origin of the misalignment and disk warping is, however, still a matter of debate.

%
% 4th paragraph
%
\cite{khetal14a} (hereafter, Paper I) derived the condition that a circumbinary disk subject 
to both the tidal torque and the torque due to the radiation emitted from two accretion disks 
around the individual supermassive black holes is unstable to the radiation-driven warping 
in the context of observed warped maser disks in active galactic nuclei (e.g., \citealt{kuo11,kho13}). 
They assumed that the binary is on a circular orbit and the irradiation luminosity is proportional 
to the mass accretion rate. There is, however, observational evidence that young binaries have 
a significant orbital eccentricity \citep{math91}. In addition, the circumbinary disk is mainly 
irradiated by two young stars in a binary in the current problem. In order to apply our previous 
model to those binaries, therefore, we need to relax the above assumptions adopted in Paper I 
and make the binary mass scale down to the stellar mass from supermassive black hole mass.

%
% 5th (last) paragraph
%
In this paper, we study the warping instability of a circumbinary disk around young binary 
stars on an eccentric orbit. In section~\ref{sec:2}, we describe the external torques acting 
on a circumbinary disk. We consider both the tidal torques originating from a time-dependent 
binary potential and the radiative torques from the binary stars. In section~\ref{sec:3}, we 
examine the evolution of a slightly tilted circumbinary disk subject to those two torques, and 
derive the warping condition and timescale of the local precession of the linear warping mode. 
Finally, section~\ref{sec:4} is devoted to summary and discussion of our scenario.

%
%%%%%%%%%%%%%%%%%%%%%%%%%%%%%
\section{External Torques acting on the circumbinary disk}
\label{sec:2}
%%%%%%%%%%%%%%%%%%%%%%%%%%%%%
%

\begin{figure}[ht!]
\begin{center}
\includegraphics[width=10cm]{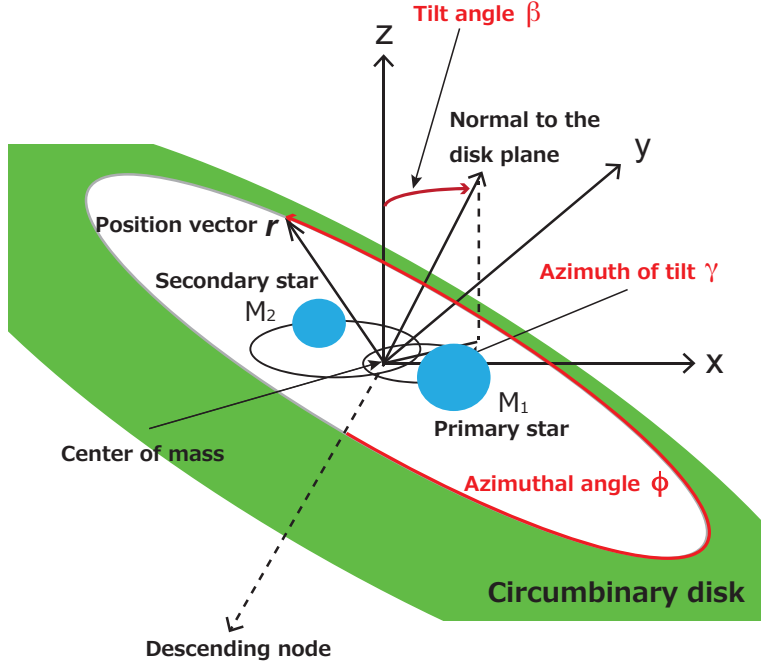}
\end{center}
\caption{
Configuration of a young binary star system composed of two stars and a circumbinary disk surrounding them. There are two angles ($\beta, \gamma$) which specify the orientation of the circumbinary disk plane with respect to the binary orbital plane ($x$-$y$ plane). The azimuthal angle ($\phi$) of an arbitrary position on the circumbinary disk is measured from the descending node.
}
\label{fig:schmatic}
\end{figure}

Let us consider the torques from the binary potential acting on the circumbinary disk 
surrounding two stars in a binary on an eccentric orbit. Figure~\ref{fig:schmatic} illustrates 
a schematic picture of the setting of our model; binary stars orbiting each other are surrounded 
by a misaligned circumbinary disk. The binary is put on the $x$-$y$ plane with its center of mass 
being at the origin in the Cartesian coordinate. The masses of the primary and secondary stars 
are represented by $M_1$ and $M_2$, respectively, and $M=M_1+M_2$. We put a circumbinary 
disk around the origin. The unit vector of specific angular momentum of the circumbinary disk is 
expressed by (e.g. \citealt{pr96})
\begin{equation}
\mbox{\boldmath $l$} = \cos\gamma\sin\beta\vec{i} + \sin\gamma\sin\beta\vec{j} +\cos\beta\vec{k},
\label{eq:damvec}
\end{equation}
where $\beta$ is the tilt angle between the circumbinary disk plane and the 
binary orbital plane, and $\gamma$ is the azimuth of tilt. 
Here, $\vec{i}$, $\vec{j}$, and $\vec{k}$ are unit vectors in the $x$, $y$, and $z$, respectively.
The position vector of the circumbinary disk can be expressed by
\begin{equation}
\mbox{\boldmath $r$}=r(\cos\phi\sin\gamma+\sin\phi\cos\gamma\cos\beta)\vec{i}
+ r(\sin\phi\sin\gamma\cos\beta-\cos\phi\cos\gamma)\vec{j} 
-r\sin\phi\sin\beta\vec{k}
\label{eq:rin}
\end{equation}
where the azimuthal angle $\phi$ is measured from the descending node. 
The only difference from the Paper~I is the position vector of each star, 
which is given by
\begin{equation}
\mbox{\boldmath $r$}_{i}=r_{i}\cos{f_i}\vec{i}+r_{i}\sin{f_i}\vec{j} \hspace{2mm}(i=1,2),
\label{eq:ri}
\end{equation}
%where {$r_{i}=\xi_{i}a$ with $\xi_1\equiv q/(1+q)$ and $\xi_2\equiv 1/(1+q)$. Here, $q=M_2/M_1$ is the binary mass ratio and $a$ is the semi-major axis of the binary.}These and other {model parameters are listed} in Table~1.
where $f_2=f_1+\pi$ is the true anomaly and $r_{i}$ is written as 
\begin{equation}
r_{i}=\xi_{i}\frac{a(1-e^2)}{1+e\cos{f_i}} 
\label{eq:rphi}
\end{equation}
with $\xi_1\equiv q/(1+q)$ and $\xi_2\equiv 1/(1+q)$. 
Here, $e$ is the binary orbital eccentricity, $q=M_2/M_1$ is the binary mass ratio, 
and $a=a_1 + a_2$ is the binary semi-major axis with $a_1\equiv\xi_1a$ and 
$a_2\equiv\xi_2a$. These and other model parameters are listed in Table~1.

%
%%%%%%%%%%%
% Table 1
%%%%%%%%%%%
%
\begin{table}[!ht]
  \caption{Model parameters}
%  Definitions for selected quantities}
%   \begin{center}
     \begin{tabular}{ll}
       \hline
       Definition & Symbol\\
       \hline
       Total stellar mass  & $M$  \\
       Primary stellar mass & $M_1$ \\
       Secondary stellar mass & $M_2$ \\
%       Schwarzschild radius & $r_{\rm{S}}=2GM/c^2$ \\
       Binary mass ratio & $q=M_2/M_1$ \\
       Mass ratio parameter{s} & $\xi_1=q/(1+q)$, $\xi_2=1/(1+q)$ \\
       Binary semi-major axis & $a$ \\ 
       Orbital eccentricity & $e$ \\
       Orbital frequency & $\Omega_{\rm{orb}}=\sqrt{GM/a^3}$ \\
       Orbital period & $P_{\rm{orb}}=2\pi/\Omega_{\rm{orb}}$ \\
       True anomaly & $f_2=f_1+\pi$ \\
%       True anomaly of the secondary & $f_2=f-\pi$ \\  
       Tilt angle & $\beta$ \\
       Azimuth of tilt & $\gamma$ \\
       Azimuthal angle & $\phi$ \\
       Shakura-Sunyaev viscosity parameter & $\alpha$ \\
       Horizontal shear viscosity & $\nu_1$ \\
       Vertical shear viscosity & $\nu_2$\\
       Ratio of vertical to horizontal shear viscosities & $\eta=\nu_2/\nu_1$\\
%       Mass-to-energy conversion efficiency & $\epsilon$ \\
       Luminosities emitted from two stars in a binary & $L_1$, $L_2$ \\
       Total luminosity & $L=L_1+L_2$ \\ 
       Binary irradiation parameter & $\zeta=(\xi_1^2L_1+\xi_2^2L_2)/L$ \\
%       Power-law index of the stellar mass-luminosity relation & $d$\\
       \hline
     \end{tabular}
   \label{tb:t1}
%   \tablecomments{{\bf More detailed explanation can be found on the text.}}
%   \end{center}
\end{table}
%%%%%%%%%%%

%
%%%%%%%
% Figure 1
%%%%%%%
%
%%%%%%
%

%
%%%%%%%%%%%%%%%%%
\subsection{Gravitational Torques}
%%%%%%%%%%%%%%%%%
%

The gravitational force {on} the unit mass at position $\vec{r}$ 
on the {circumbinary disk} can be written by
\begin{eqnarray}
\vec{F}_{\rm{grav}}=-\sum_{i=1}^{2}\frac{GM_i}{|\vec{r}-\vec{r}_i|^3}(\vec{r}-\vec{r}_i)
\end{eqnarray}
The corresponding torque is given by
\begin{eqnarray}
\vec{t}_{\rm{grav}}=\vec{r}\times\vec{F}_{\rm{grav}}=\sum_{i=1}^2\frac{GM_i}{|\vec{r}-\vec{r}_i|^3}(\vec{r}\times\vec{r}_i)
\end{eqnarray}
%%%
We consider the tidal warping/precession with timescales much longer than 
local rotation period of the circumbinary disk. This allows us to use the torque 
averaged in the azimuthal direction and over the orbital period:
\begin{eqnarray}
\langle\vec{T_{\rm{grav}}}\rangle
%&=&
&
\approx
&
\frac{1}{4\pi^2}\int_0^{2\pi}
\int_0^{2\pi}
\vec{t}_{\rm{grav}}
\,d\phi
d(\Omega_{\rm{orb}}t)
=\frac{3}{8}\xi_1\xi_2
\left(\frac{a}{r}\right)^2
\frac{GM}{r}
\nonumber \\
&\times&
\Biggr[
(1-e^2)
\sin\gamma\sin2\beta\vec{i}
-
(1+4e^2)
\cos\gamma\sin2\beta\vec{j}
+5e^2\sin2\gamma\sin^2\beta\vec{k}
\Biggr],
\label{eq:tgrav}
\end{eqnarray}
where $\Omega_{\rm{orb}}=\sqrt{GM/a^3}$ is the angular frequency of mean binary motion. 
Here, we used for the integration the following relationship:
\begin{eqnarray}
d(\Omega_{\rm{orb}}{t})=\frac{(1-e^2)^{3/2}}{(1+e\cos{f}_i)^{2}}df
\end{eqnarray}
and the approximations:
\begin{eqnarray}
|\vec{r}-\vec{r}_i|^{-3}&\approx&r^{-3}\left[1+3\frac{\vec{r}\cdot\vec{r}_i}{r^2}+\mathcal{O}((r_i/r)^2)\right],
\nonumber \\
(1+e\cos{f}_i)^{-4}&\approx&1-4e\cos{f}_i+10e^2\cos^2{f}_i+\mathcal{O}(e^3).
\end{eqnarray}
Note that equation~(\ref{eq:tgrav}) is derived in the same manner as in equation~(7) of \cite{khetal14b}.

For a small tilt angle $\beta\ll1$, equation~(\ref{eq:tgrav}) is reduced to 
\begin{eqnarray}
\langle\vec{T}_{\rm{grav}}\rangle
\approx
\frac{3}{4}
\xi_1\xi_2
\left(\frac{a}{r}\right)^2
\frac{GM}{r}
\Biggr[
(1-e^2)
l_y
\vec{i}
-
(1+4e^2)
l_x
\vec{j}
\Biggr],
\label{eq:tgrav2}
\end{eqnarray}
where $l_x$ and $l_y$ can be written from equation~(\ref{eq:damvec}) as $l_x=\beta\cos\gamma$ 
and $l_y=\beta\sin\gamma$. The magnitude of the specific tidal torque with $\beta\ll1$ is then given by 
\begin{eqnarray}
|\langle\vec{T}_{\rm{grav}}\rangle|
=
\frac{3}{4}
\xi_1\xi_2
\left(\frac{a}{r}\right)^2
\frac{GM}{r}
\beta\sqrt{5e^2(3e^2+2)\cos^2\gamma+(e^2-1)^2}.
\label{eq:tidmag}
\end{eqnarray}
 
The tidal torque tends to align the tilted circumbinary disk with the orbital plane (c.f. \citealt{bate00}). 
The tidal alignment timescale for $\beta\ll1$ is given by
\begin{eqnarray}
\tau_{\rm{tid}}=
\frac{|\vec{J}|\sin\beta}{|\langle\vec{T}_{\rm{grav}}\rangle|}
\approx\frac{8}{3\pi}
\frac{1}{\sqrt{5e^2(3e^2+2)\cos^2\gamma+(e^2-1)^2}}
\left(\frac{1/4}{\xi_1\xi_2}\right)
\left(\frac{r}{a}\right)^{7/2}
P_{\rm{orb}},
\label{eq:ttid}
\end{eqnarray}
where 
$\vec{J}\equiv{r}^2\Omega\vec{l}$ with the disk angular frequency 
$\Omega=\sqrt{GM/r^3}$ and $P_{\rm{orb}}=2\pi/\Omega_{\rm{orb}}$ 
are the specific angular momentum and binary orbital period, respectively.
Equation~(\ref{eq:ttid}) is reduced to equation~(8) of Paper I for $e=0$.
Note that the tidal alignment timescale depends on the azimuth of tilt for $e\neq0$. 
Since the inner edge of the circumbinary disk is estimated to be $\sim2a$ \citep{al94}, 
the tidal alignment timescale is longer than the binary orbital period.

%
%%%%%%%%%%%%%%%%%%%%%%%%%%%%%%
\subsection{Radiative Torques}
%%%%%%%%%%%%%%%%%%%%%%%%%%%%%%
%

%
% 1st paragraph
%
The circumbinary disk around young binary stars can be mainly illuminated by light 
emitted from each star. The re-radiation from the circumbinary-disk surface, which 
absorbs photons emitted from these stars, causes a reaction force. This is the origin 
of the radiative torques. Here, the central two stars are can be regarded as point 
irradiation sources, because each star is much smaller than the size of circumbinary 
disk. Circumstellar disks in binary systems have been observed by many authors in 
the past (e.g. \citealt{maya10}). These circumstellar disks could be other irradiation 
sources. We estimate their luminosities as follows: the maximum luminosity of each 
disk is given by $L_{\rm{disk}}=GM\dot{M}_{\rm{crit}}/R_{*}\sim7\times10^{-2}\,L_\odot(M/M_\odot)^2
(R_*/R_\odot)^{-1}$ with $\dot{M}_{\rm{crit}}=L_{\rm{Edd}}/c^2$, where the Eddington 
luminosity is given by $L_{\rm{Edd}}\simeq1.3\times10^{38}\,(M/M_\odot)\,{\rm erg\,s^{-1}}$. 
Since it is substantially smaller than the stellar luminosity, the effect of the circumstellar-disk 
irradiation is negligible.

Since the surface element on the circumbinary disk is given in the polar 
coordinates by
\begin{eqnarray}
\vec{dS}
=\frac{\partial\vec{r}}{\partial{r}}\times\frac{\partial\vec{r}}{\partial\phi}\,\,drd\phi
=
\left[
\vec{l}-\vec{r}
\left(
-\frac{\partial\beta}{\partial{r}}\sin\phi
+\frac{\partial\gamma}{\partial{r}}\cos\phi\sin\beta
\right)
\right]
\,rdrd\phi,
\end{eqnarray}
the radiative flux at $d\vec{S}$ is given by
\begin{eqnarray}
dL
=\frac{1}{4\pi}\sum_{i=1}^{2}\frac{L_i}{|\vec{r}-\vec{r}_i|^2}
\frac{|(\vec{r}-\vec{r}_i)\cdot{d}\vec{S}|}{|\vec{r}-\vec{r}_i|},
\end{eqnarray}
where $L$ is the sum of luminosities of the radiation emitted from the primary star, $L_1$,
and that from the secondary star, $L_2$. Here, we assume that the surface element is not shadowed by other interior parts of the circumbinary disk. If we ignore limb darkening, the force acting on the disk surface by the radiation reaction has the magnitude of $(2/3)(dL/c)$ and is antiparallel to the local disk normal \citep{pr96}.
The total radiative force on $d\vec{S}$ can then be written by
\begin{eqnarray}
d\vec{F}_{\rm{rad}}
=
\frac{1}{6\pi{c}}
\sum_{i=1}^{2}L_i
\frac{|(\vec{r}-\vec{r}_i)\cdot
d\vec{S}|}{|\vec{r}-\vec{r}_i|^3}
\frac{d\vec{S}}{|d\vec{S}|}.
\end{eqnarray}
Consequently, the total radiative torque acting on a ring of radial width $dr$ is given by
\begin{eqnarray}
d\vec{T}_{\rm{rad}}
=
\oint\vec{r}\times{d\vec{F}_{\rm{rad}}}
%=\frac{1}{6\pi{c}}\oint\sum_{i=1}^{2}\Biggr[L_i\frac{|(\vec{r}-\vec{r}_i)\cdotd\vec{S}|}{|\vec{r}-\vec{r}_i|^3}\Biggr]\frac{\vec{r}\times{d}\vec{S}}{|d\vec{S}|}
%\nonumber \\
%&\approx&
\approx
\frac{L}{6\pi{c}}
\frac{1}{r^3}
\oint
|
\vec{r}\cdot{d}\vec{S}
|
\frac{\vec{r}\times{d}\vec{S}}{|d\vec{S}|}
+
\frac{1}{2\pi{c}}
\frac{1}{r^5}
\oint
\sum_{i=1}^{2}
L_i
|
(\vec{r}-\vec{r}_i)\cdot{d}\vec{S}
|
(\vec{r}\cdot\vec{r}_{i})
\frac{\vec{r}\times{d}\vec{S}}{|d\vec{S}|},
%\nonumber \\
\label{eq:radtorque}
\end{eqnarray}
where  $\oint|(\vec{r}-\vec{r}_i)\cdot{d}\vec{S}|(\vec{r}\times{d}\vec{S})/|d\vec{S}|=\oint|\vec{r}\cdot{d}\vec{S}|(\vec{r}\times{d}\vec{S})/|d\vec{S}|$ holds for $\beta\ll1$ and ${r}\partial{\beta}/\partial{r}\ll1$, 
and the first term, which we call $d\vec{T}_0$, of the right-hand side of equation~(\ref{eq:radtorque}) corresponds to equation (2.15) of \cite{pr96}:
\begin{eqnarray}
d\vec{T}_0=\frac{L}{6c}\left(r\frac{\partial{l_y}}{\partial{r}}\vec{i}-r\frac{\partial{l_x}}{\partial{r}}\vec{j}
\right)dr
\label{eq:dt0}
\end{eqnarray}
and the second term, which we call $d\vec{T}_{\rm{orb}}$, is originated from the orbital motion of the binary.

%
% 2nd paragraph
%
Here, we consider the radiation-driven warping/precession with timescales much longer than 
the orbital period, as in the case of tidally driven warping/precession.
The orbit-average of the torque $d\vec{T}_{\rm{rad}}$ is then given by
\begin{eqnarray}
\langle
d\vec{T}_{\rm{rad}}
\rangle
&=&
\frac{1}{2\pi}\int_{0}^{2\pi}
d\vec{T}_{\rm{rad}}\,d(\Omega_{\rm{orb}}t)
\approx
{d}\vec{T}_0+\frac{1}{2\pi}\int_{0}^{2\pi}d\vec{T}_{\rm{orb}}\,d(\Omega_{\rm{orb}}t)
\nonumber \\
&=&
\frac{L}{6c}
\Biggr\{
\Biggr(
-\frac{3}{2}\zeta(1-e^2)\left(\frac{a}{r}\right)^2
l_y
+r\left[1-\frac{3}{8}\zeta(4+e^2)\left(\frac{a}{r}\right)^2
\right]\frac{\partial{l_y}}{\partial{r}}
\Biggr)\vec{i}
\nonumber \\
&+&
\Biggr(
\frac{3}{2}\zeta(1+4e^2)\left(\frac{a}{r}\right)^2
{l_x}
-r\left[1-\frac{3}{8}\zeta(4+11e^2)\left(\frac{a}{r}\right)^2
\right]\frac{\partial{l_x}}{\partial{r}}
\Biggr)\vec{j}
\Biggr\}dr,
\label{eq:radt}
\end{eqnarray}
where $\zeta\equiv(\xi_1^2L_1+\xi_2^2L_2)/L$ is a binary irradiation parameter ($\zeta<1$ by definition), 
and $\langle{d\vec{T}_{\rm{rad}}}\rangle$ is reduced to $d\vec{T}_0$ for $r\gg{a}$.

%
% 3rd paragraph
%
From equation~(\ref{eq:radt}), the specific radiative torque averaged over azimuthal angle and orbital phase 
is given by
\begin{eqnarray}
 \langle\vec{T}_{\rm{rad}}\rangle
 &=&
\frac{1}{2\pi{r}\Sigma}\frac{\langle{d}\vec{T}_{\rm{rad}}\rangle}{dr}
=
|\vec{J}|\frac{\Gamma}{r}
\Biggr\{
\Biggr(
-\frac{3}{2}\zeta(1-e^2)\left(\frac{a}{r}\right)^2
{l_y}
+r\left[1-\frac{3}{8}\zeta(4+e^2)\left(\frac{a}{r}\right)^2
\right]\frac{\partial{l_y}}{\partial{r}}
\Biggr)\vec{i}
\nonumber \\
&+&
\Biggr(
\frac{3}{2}\zeta(1+4e^2)\left(\frac{a}{r}\right)^2l_x
-r\left[1-\frac{3}{8}\zeta(4+11e^2)\left(\frac{a}{r}\right)^2
\right]\frac{\partial{l_x}}{\partial{r}}
\Biggr)\vec{j}
\Biggr\},
\end{eqnarray}
where $\Gamma=L/(12\pi\Sigma{r^2}\Omega{c})$ and $\Sigma$ are the growth speed of a warping mode 
induced by the radiative torque and disk surface density, respectively. The growth timescale of the warping 
mode for an optically thick gas disk can be estimated to be
\begin{eqnarray}
\tau_{\rm{rad}}=\frac{r}{\Gamma}=6GM\Sigma\left(\frac{c}{L}\right)\left(\frac{r}{a}\right)^{3/2}
P_{{\rm orb}}=6.2\times10^3
\left(\frac{r}{a}\right)^{3/2}
\left(\frac{\Sigma}{1\,\rm{g\,cm^{-2}}}\right)
\left(\frac{M}{\rm M_\odot}\right)
\left(\frac{L}{\rm L_\odot}\right)^{-1}
P_{\rm orb}.
\label{eq:trad}
\end{eqnarray}
Since it is clear that the growth timescale for $r>a$ is much longer that the orbital period, 
our assumption for the orbit-averaged radiative torque is ensured.

%
%%%%%%%%%%%%%%%%%%%%%%%%%
\section{Tilt angle evolution of circumbinary disks}
\label{sec:3}
%%%%%%%%%%%%%%%%%%%%%%%%%
%
% 1st paragraph
%
In this section, we examine the response of the circumbinary disk to the external forces 
for $\beta\ll1$ case. The evolution equation for disk tilt is given by\citep{pr96}
\begin{equation}
\frac{\partial\vec{l}}{\partial{t}}
+
\left[
v_r-\nu_1\frac{\Omega^{'}}{\Omega}
-\frac{1}{2}\nu_2\frac{(r^3\Omega\Sigma)^{'}}{r^3\Omega\Sigma}
\right]\frac{\partial\vec{l}}{\partial{r}}
=\frac{\partial}{\partial{r}}\left(\frac{1}{2}\nu_2\frac{\partial{\vec{l}}}{\partial{r}}\right)
+\frac{1}{2}\nu_2\left|\frac{\partial\vec{l}}{\partial{r}}\right|^2\vec{l}+\vec{T}_{\rm{ex}},
\label{eq:dtilt}
\end{equation}
where $\nu_1$ and $\nu_2$ are respectively the horizontal and vertical shear viscosities, 
the latter of which tends to reduce disk tilt, and $\vec{T}_{\rm ex}$ is the term to which the 
external torques contribute. For a geometrically thin disk, $\nu_1$ is approximately given by 
$\nu_1\approx\alpha{c_{\rm{s}}^2}/\Omega=\alpha({\rm{R_{g}}}/\mu)T/\Omega$ with the 
Shakura-Sunyaev viscosity parameter $\alpha$ \citep{ss73}, the isothermal sound speed $c_{\rm{s}}$, 
the temperature $T$, the gas constant ${\rm{R_g}}$, and the molecular weight $\mu$.
The primes indicate differentiation with respect to $r$. 
We adopt for the circumbinary disk structure the following assumptions that $v_r=\nu_1\Omega^{'}/\Omega$, 
\begin{eqnarray}
\Sigma(r)&=&\Sigma_0\left(\frac{r}{r_0}\right)^n \hspace{3mm}(n<0),
\label{eq:sigma}
\\
T(r)&=&T_0\left(\frac{r}{r_0}\right)^s \hspace{4mm}(s<0),
\label{eq:temp}
\end{eqnarray}
where $n$ and $s$ are a constant, and $r_0$, $\Sigma_0$, and $T_0$ are the fiducial radius, 
fiducial surface density, and fiducial temperature, respectively. Equation (\ref{eq:dtilt}) can be 
then reduced to
\begin{eqnarray}
\frac{\partial\vec{l}}{\partial{t}}=\frac{1}{2}\nu_2\frac{\partial^2\vec{l}}{\partial{r}^2}
+\frac{1}{2}\left(n+s+3\right)\frac{\nu_2}{r}\frac{\partial\vec{l}}{\partial{r}}
+\vec{T}_{\rm{ex}},
\label{eq:tiltevo}
\end{eqnarray}
where $\vec{l}\cdot\partial\vec{l}/\partial{r}=0$ is used. 
Here, $\nu_2\equiv\eta\nu_1\approx\eta\alpha({\rm{R_g}}/\mu)(T(r)/\Omega)$ 
with the ratio of vertical to horizontal viscosities: $\eta=\nu_2/\nu_1$ and $\vec{T}_{\rm{ex}}$ is written as
\begin{eqnarray}
\vec{T}_{\rm{ex}}
&=&
(\langle\vec{T}_{\rm{grav}}\rangle + \langle\vec{T}_{\rm{rad}}\rangle)/|\vec{J}|
\nonumber \\
&=&
\Biggr\{
\frac{3}{2}\left(\frac{a}{r}\right)^2(1-e^2)
\left[
\frac{1}{2}
\xi_1\xi_2
\Omega
-\frac{\zeta}{\tau_{\rm{rad}}}
\right]l_y
+\Gamma\left[1-\frac{3}{8}\zeta\left(\frac{a}{r}\right)^2(4+e^2)
\right]\frac{\partial{l_y}}{\partial{r}}
\Biggr\}\vec{i}
\nonumber \\
&-&
\Biggr\{
\frac{3}{2}\left(\frac{a}{r}\right)^2(1+4e^2)
\left[
\frac{1}{2}
\xi_1\xi_2
\Omega
-\frac{\zeta}{\tau_{\rm{rad}}}
\right]l_x
+\Gamma\left[1-\frac{3}{8}\zeta\left(\frac{a}{r}\right)^2(4+11e^2)
\right]\frac{\partial{l_x}}{\partial{r}}
\Biggr\}\vec{j}.
\end{eqnarray}

%
% 2nd paragraph
%
We look for solutions of equation (\ref{eq:tiltevo}) of the form $l_x$, $l_y$$\propto\exp{i}(\omega{t}+kr)$ with $kr\ll1$.
Replacing $\partial/\partial{t}$ with $i\omega$, $\partial/\partial{r}$ with $ik$, and $\partial^2/\partial{r^2}$ with $-k^2$, 
we have the following set of linearized equations:
\begin{eqnarray}
\left[
\begin{array}{cccc}
i\omega + \nu_2k^2/2 -(ik/2)(n+s+3)(\nu_2/r) && -(\mathcal{A}+ik\mathcal{B}) \\
(\mathcal{C}+ik\mathcal{D})  && i\omega+\nu_2{k^2}/2 -(ik/2)(n+s+3)(\nu_2/r) \\
\end{array}
\right]
\left(
\begin{array}{cc}
l_x \\
l_y  \\
\end{array}
\right)
=0,
\label{eq:eigen}
\end{eqnarray}
where
\begin{eqnarray}
\mathcal{A}&=&
(1-e^2)\Omega_{\rm{p,circ}},
\nonumber \\
\mathcal{B}&=&\Gamma\left[1-\frac{3}{8}\zeta\left(\frac{a}{r}\right)^2(4+e^2)
\right],
\nonumber \\
\mathcal{C}&=&(1+4e^2)\Omega_{\rm{p,circ}},
\nonumber \\
\mathcal{D}&=&\Gamma\left[1-\frac{3}{8}\zeta\left(\frac{a}{r}\right)^2(4+11e^2)
\right],
%\nonumber 
\end{eqnarray}
and, $\Omega_{\rm{p},\rm{circ}}$ represents the magnitude of the local precession frequency of the liner warping mode for the case of the standard disk model and $e=0$ (see equation~(35) of Paper~I):
\begin{eqnarray}
\Omega_{\rm{p},\rm{circ}}
&=&
\frac{3}{2}\left(\frac{a}{r}\right)^2\left[
\frac{1}{2}
\xi_1\xi_2
\Omega
-\frac{\zeta}{\tau_{\rm{rad}}}
\right].
\end{eqnarray}
%This approximation is held as long as the outer radius of the circumbinary disk is constrained by the dust sublimation radius (see equation~(36) of Paper~I in detail).
%
The determinant of the coefficient matrix on the left hand side of equation~(\ref{eq:eigen}) 
must {vanish} because of $\vec{l}\neq0$. The local dispersion relation is then obtained as
\begin{eqnarray}
\omega
=
i
\left[
\frac{\nu_2k^2}{2}\pm\frac{1}{\sqrt{2}}(\sqrt{X^2+Y^2}-X)^{1/2}
\right]
\pm
\frac{1}{\sqrt{2}}(\sqrt{X^2+Y^2}+X)^{1/2},
\label{eq:omega}
\end{eqnarray}
where 
\begin{eqnarray}
X&=&\mathcal{AC}-k^2\mathcal{BD}\approx(1+3e^2)\Omega_{\rm{p,circ}}^2-(k\Gamma)^2\left[1-3\zeta\left(1+\frac{3}{2}e^2\right)\left(\frac{a}{r}\right)^2\right],
\nonumber \\
Y&=&\mathcal{AD+BC}\approx-2k\Gamma\Omega_{\rm{p,circ}}\left[\left(1+\frac{3}{2}e^2\right)-\frac{3}{2}\zeta(1+3e^2)\left(\frac{a}{r}\right)^2\right],
\nonumber \\
X^2+Y^2
&\approx&
\Biggr\{
(1+3e^2)\Omega_{\rm{p,circ}}^2+
(k\Gamma)^2\left[1-3\zeta\left(1+\frac{3}{2}e^2\right)\left(\frac{a}{r}\right)^2\right]
\Biggr\}^2.
\end{eqnarray}
Here, we adopt the approximation that the terms proportional to $(r/a)^4$ 
or $e^4$ are negligible in comparison with the other terms. 
The dispersion relation is then rewritten in the following simple form: 
\begin{eqnarray}
\omega
=
i\Biggr\{
\frac{\nu_2k^2}{2}\pm
{k}\Gamma\left[1-3\zeta\left(1+\frac{3}{2}e^2\right)\left(\frac{a}{r}\right)^2\right]^{1/2}
\Biggr\}
\pm
\sqrt{1+3e^2}
\Omega_{\rm{p,circ}}
+\frac{1}{2}\frac{{k}\nu_2}{r}(n+s+3)
\label{eq:omega}
\end{eqnarray}
The imaginary part of $\omega$ corresponds to the excitation or damping of oscillation, whereas the real part 
provides the local precession frequency due to the external torques.

%
%%%%%%%
% Figure 2
%%%%%%%
%
\begin{figure}[ht!]
\resizebox{\hsize}{!}{
\includegraphics{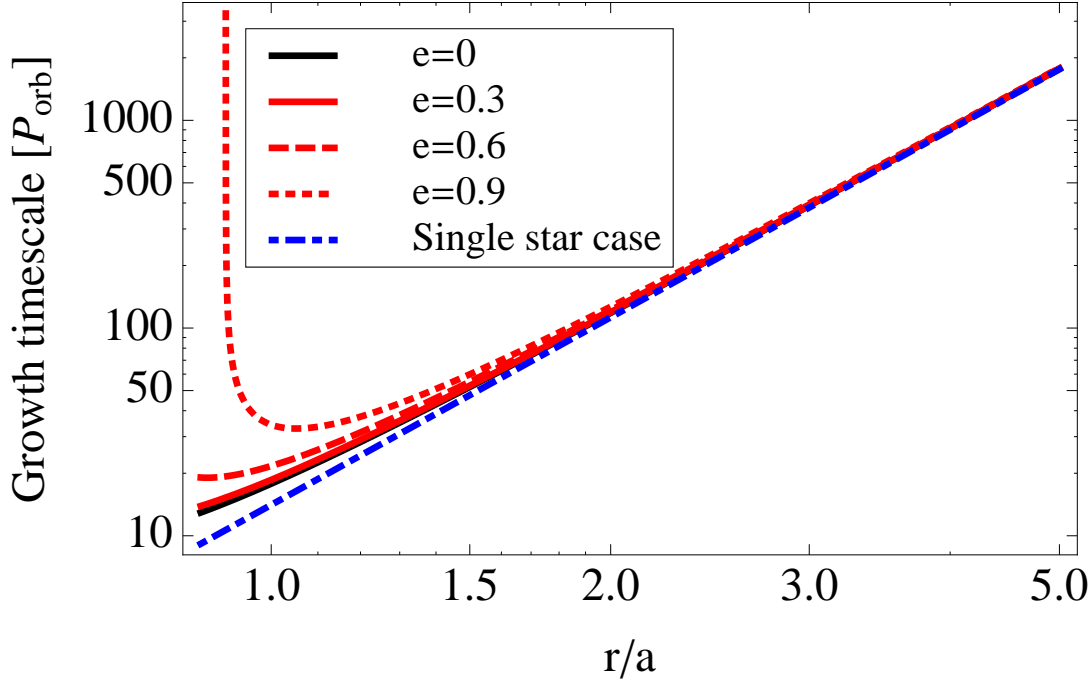}
}
\caption{
Growth timescale of the radiation-driven warping of a circumbinary disk. 
Here we adopted that $\zeta=1/4\,(q=1)$, $\Sigma=1\,{\rm{g\,cm^{-2}}}$, $L=10^{4}\,{L_\odot}$, 
$M=30\,{M}_\odot$, and $a=1\,{\rm{AU}}$. Each growth timescale is normalized by $P_{\rm{orb}}\simeq67\,{\rm{days}}$. The black solid, red solid, red dashed, and red dotted lines show the growth timescales for the orbital eccentricity $e=0.0$, $0.3$, $0.6$, and $0.9$, respectively. The blue dot-dashed line corresponds to that of the single star case.
}
\label{fig:tgrowth}
\end{figure}
%%%%%%

%
% 3rd paragraph
%
In order for the perturbation to grow, $\rm{Im}(\omega)$ must be negative. 
The growth condition is given by
\begin{eqnarray}
0<k<\frac{2\Gamma}{\nu_2}\left[1-3\zeta\left(1+\frac{3}{2}e^2\right)\left(\frac{a}{r}\right)^2\right]^{1/2}.
\label{eq:waven}
\end{eqnarray}
In terms of $\Gamma_{\rm{bin}}\equiv\Gamma[1-3\zeta(1+3e^2/2)(a/r)^2]^{1/2}$, the growth timescale of the warping mode induced 
by the radiative torques in the binary system is given by
\begin{eqnarray}
\tau_{\rm{rad,bin}}=\frac{r}{\Gamma_{\rm{bin}}}=\tau_{\rm{rad}}\left[1-3\zeta\left(1+\frac{3}{2}e^2\right)
\left(\frac{a}{r}\right)^2\right]^{-1/2}.
\label{eq:radt_bin}
\end{eqnarray}
Figure~\ref{fig:tgrowth} shows the dependence of $\tau_{\rm{rad,bin}}$ on the orbital eccentricity 
and $r/a$ in a circumbinary disk with $\zeta=1/4\,(q=1)$, $\Sigma=1\,{\rm{g\,cm^{-2}}}$, $L=10^{4}\,{L_\odot}$, $M=30\,{M}_\odot$, and $a=1\,{\rm{AU}}$. Each growth timescale is normalized by 
$P_{\rm{orb}}\simeq67\,{\rm{days}}$. The growth timescale is longer than that of circular binary case in the range of $r/a<3$. This is because the averaged incident flux normal to the disk surface per binary orbit is lower than that of the circular binary case. Note that $\tau_{\rm{bin,rad}}$ is approximately equal to $\tau_{\rm{rad}}$ for $r/a\ge3$. We therefore treat $\tau_{\rm{rad}}$ as the growth timescale of the binary star case in what follows.

%
% 4th paragraph
%
We focus our attention on a perturbation with $\lambda\le{r}$, 
where $\lambda=2\pi/k$ is the radial wavelength of the perturbation. 
The condition that the circumbinary disk is unstable to the warping mode 
can be, then, rewritten as
\begin{eqnarray}
r\le
\frac{1}{12\pi^2\eta\alpha\Sigma{c_{\rm{s}}}^2}\left(\frac{L}{c}\right)\left[1-3\zeta\left(1+\frac{3}{2}e^2\right)\left(\frac{a}{r}\right)^{2}\right]^{1/2}.
\label{eq:cond}
\end{eqnarray}
Substituting equations~(\ref{eq:sigma}) and (\ref{eq:temp}) into the above equation, we obtain 
\begin{eqnarray}
\frac{r}{r_0}
\left\{ \begin{array}{ll}
\ge (r_{\rm{warp}}/r_0)\left[1-3\zeta\left(1+\frac{3}{2}e^2\right)\left(\frac{a}{r}\right)^2\right]^{1/(2(n+s+1))}
& (n+s+1<0) \\
 \le (r_{\rm{warp}}/r_0)\left[1-3\zeta\left(1+\frac{3}{2}e^2\right)\left(\frac{a}{r}\right)^2\right]^{1/(2(n+s+1))} & (n+s+1>0),
\end{array} \right.
\label{eq:rwbin}
\end{eqnarray}
where $r_{\rm{warp}}$ shows the marginally stable warping radius for a single star case:
\begin{eqnarray}
\frac{r_{\rm warp}}{r_0}
=\left[
12\pi^2\eta\alpha
r_0\Sigma_0
\frac{\rm{R_g}}{\mu}T_0
\frac{c}{L}
\right]^{-1/(n+s+1)}
\approx
\left[
6\pi^2\frac{1}{\alpha}
r_0\Sigma_0
\frac{\rm{R_g}}{\mu}T_0
\frac{c}{L}
\right]^{-1/(n+s+1)},
\label{eq:rw}
\end{eqnarray}
where in deriving the second equation, we have used the relationship between $\eta$ 
and $\alpha$: $\eta=2(1+7\alpha^2)/(\alpha^2(4+\alpha^2))\approx1/(2\alpha^2)$ 
for $\alpha\ll1$ \citep{og99}. In the case of $n+s+1=0$, no unstable solution exists except 
for very special combination of parameters.

%
% 5th paragraph
%
The radiative torques work only for a region of optically thick to irradiation emitted from the central stars.
In order for the circumbinary disk to be optically thick, the surface density must be higher than 
$\Sigma_{\rm{min}}\simeq1\,\rm{g\,cm^{-2}}$, where the electron scattering opacity is assumed for simplicity. 
This condition is rewritten as 
\begin{eqnarray}
\frac{r}{r_0}\le\left(\frac{\Sigma_{\rm{min}}}{\Sigma_0}\right)^{1/n},
\end{eqnarray}
using equation~(\ref{eq:sigma}).
The equality is held at the radius where the circumbinary disk changes 
from optically thick to optically thin:
\begin{eqnarray}
\frac{r_{\rm{op}}}{r_0}=\left(\frac{\Sigma_{\rm{min}}}{\Sigma_0}\right)^{1/n}.
\label{eq:rop}
\end{eqnarray}
In order for radiation-driven warping to be a possible mechanism for disk warping, 
the marginally stable warping radius must be less than the outer radius of the optically thick region, 
$r_{\rm{op}}$. This gives the upper (lower) limit of $\Sigma_0$ for $n+s+1<0\,(n+s+1>0)$ as follows:
\begin{eqnarray}
\Sigma_0
\left\{ \begin{array}{ll}
\le \Sigma_{\rm{crit}}
\left[1-3\zeta\left(1+\frac{3}{2}e^2\right)\left(\frac{a}{r}\right)^2\right]^{-n/(2(s+1))}
& (n+s+1<0) \\
\ge \Sigma_{\rm{crit}}
\left[1-3\zeta\left(1+\frac{3}{2}e^2\right)\left(\frac{a}{r}\right)^2\right]^{-n/(2(s+1))}
 & (n+s+1>0),
\end{array} \right.
\end{eqnarray}
where $\Sigma_{\rm{crit}}$ is the critical surface density given by
\begin{eqnarray}
\Sigma_{\rm{crit}}=\Sigma_{\rm{min}}^{(n+s+1)/(s+1)}
\left[
12\pi^2\eta\alpha
r_0
\frac{\rm{R_g}}{\mu}T_0
\frac{c}{L}
\right]^{n/(s+1)}.
\label{eq:scrit}
\end{eqnarray}

%
%%%%%%%
% Figure 3
%%%%%%%
%
\begin{figure}[ht!]
\begin{center}
\resizebox{\hsize}{!}
{
\includegraphics[]{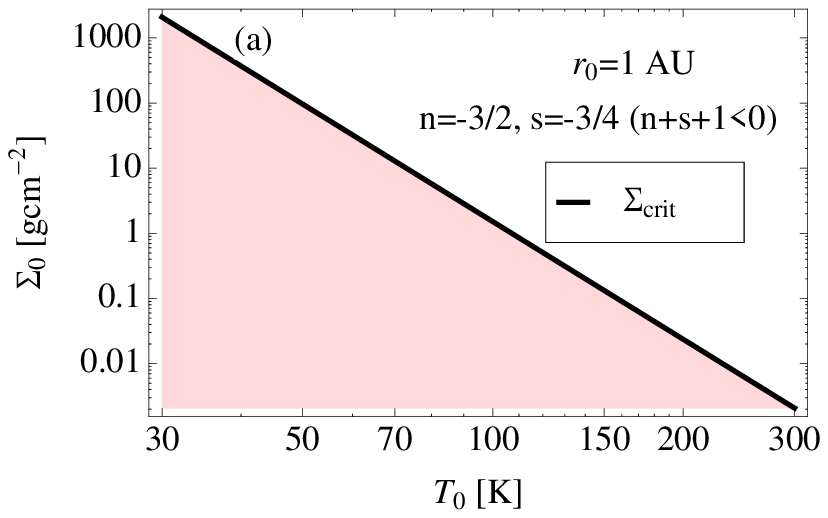}
\includegraphics[]{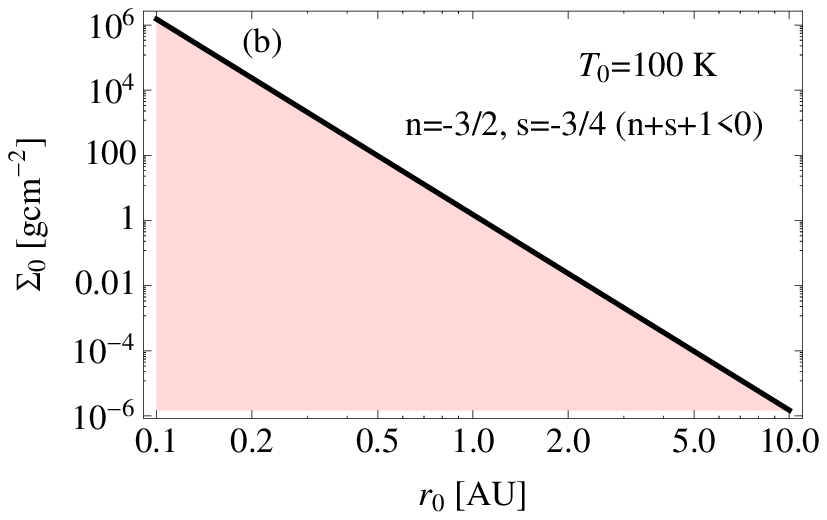}
}
\resizebox{\hsize}{!}
{
\includegraphics[]{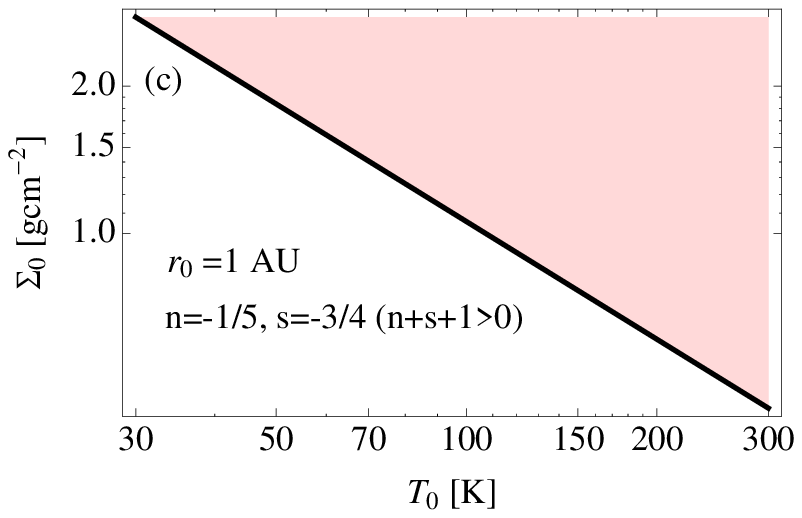}
\includegraphics[]{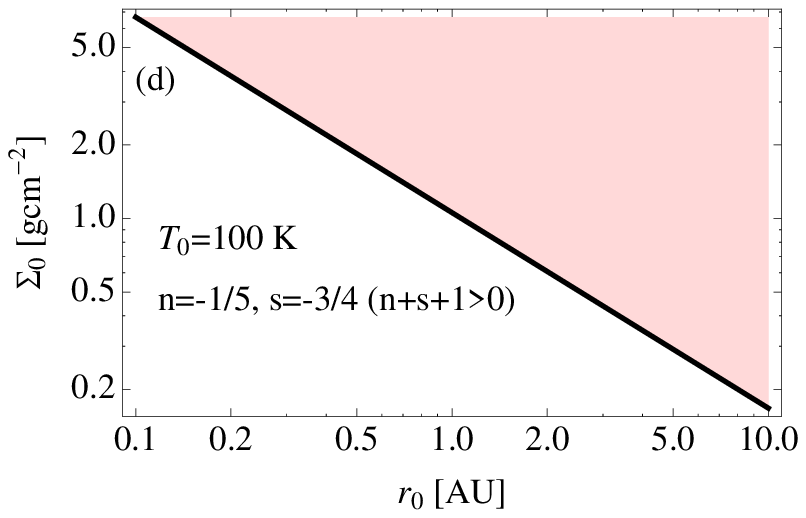}
}\\
\end{center}
\caption{
Possible range in $\Sigma_0$ (shaded area), for the circumbinary disk to be unstable to warping modes, 
as a function of $T_0$ and $r_0$ in the case of $r\gg{a}$. The black line shows a critical surface density, 
which is given by equation~(\ref{eq:scrit}), for $\Sigma_{\rm{min}}=1\,\rm{g\,cm^{-2}}$, $\alpha=0.01$, and 
$L=10^4L_\odot$. The values of $n$ and $s$ are annotated in each panel.
}
\label{fig:sfid}
\end{figure}
%%%%%%
%

%
% 6th paragraph
%
Figure~\ref{fig:sfid} shows the possible range in $\Sigma_0$, for the circumbinary disk to be unstable 
to warping modes, as a function of $T_0$ and $r_0$. The upper two panels give the range for $n+s+1<0$ 
case, whereas the lower two panels for $n+s+1>0$ case. From the figure, we can express $\Sigma_0$ as 
\begin{eqnarray}
\Sigma_0=f\Sigma_{\rm{crit}}\left[1-3\zeta\left(\frac{a}{r}\right)^2\left(1+\frac{3}{2}e^2\right)\right]^{-n/(2(s+1))}
\label{eq:sfid}
\end{eqnarray}
with a factor $f$ in the range $0<f\le1$ for $n+s+1<0$ and $f>1$ for $n+s+1 >0$. 
Substituting equation~(\ref{eq:sfid}) into equation~(\ref{eq:rw}) with equation~(\ref{eq:rwbin}),
we can obtain the marginally stable warping radius as 
\begin{eqnarray}
r_{\rm{warp,bin}}
=
r_{\rm{warp}}
\left[1-3\zeta\left(1+\frac{3}{2}e^2\right)\left(\frac{a}{r}\right)^2\right]^{1/(2(s+1))},
\label{eq:rwb}
\end{eqnarray}
where $r_{\rm{warp}}$ can be rewritten as
\begin{eqnarray}
r_{\rm{warp}}
&=&
f^{-1/(n+s+1)}
\left[
6\pi^2
\frac{1}{\alpha}
r_0
\Sigma_{\rm{min}}
\frac{\rm{R_g}}{\mu}T_0
\frac{c}{L}
\right]^{-1/(s+1)}
r_0
\simeq
10^{-4/(s+1)}\,[\rm{AU}]
\left(\frac{1}{\it f}\right)^{1/(n+s+1)}
\nonumber \\
&\times&
\left[
\left(\frac{\alpha}{0.01}\right)
\left(\frac{r_0}{1\,\rm{AU}}\right)^{s}
\left(\frac{\Sigma_{\rm{min}}}{1\rm{g\,cm^{-2}}}\right)^{-1}
\left(\frac{T_0}{100\,\rm{K}}\right)^{-1}
\left(\frac{L}{L_\odot}\right)
\right]^{1/(s+1)}.
\label{eq:rws}
\end{eqnarray}
Adopting for equation~(\ref{eq:rws}) $s=-3/4$, which corresponds to the power law index 
of the radial temperature profile of an optically thick region in the standard disk \citep{ss73}, 
the marginally stable warping radius is very sensitive to the values of $\alpha$, $r_0$, 
$\Sigma_{\rm{min}}$, $T_0$, and $L$. It is clear from equations~(\ref{eq:rwb}) and 
(\ref{eq:rws}) that a circumbinary disk around young binary stars with ${M}\sim1\,M_\odot$ 
and $L\sim1\,L_\odot$ is stable for radiation driven warping, because the marginally stable 
warping radius is of the order of $10^{-16}\,\rm{AU}$ for $s=-3/4$ and of $10^{-8}\,\rm{AU}$ 
even for $s=-1/2$. This shows that the circumbinary disks of a classical T Tauri star system 
such as GG Tau \citep{dut94} and UY Aurigae \citep{duv98} are not warped by radiation-driven 
warping instability. 

%
%%%%%%%
% Figure 3
%%%%%%%
%
\begin{figure}[ht!]
\begin{center}
\resizebox{\hsize}{!}
{
\includegraphics[]{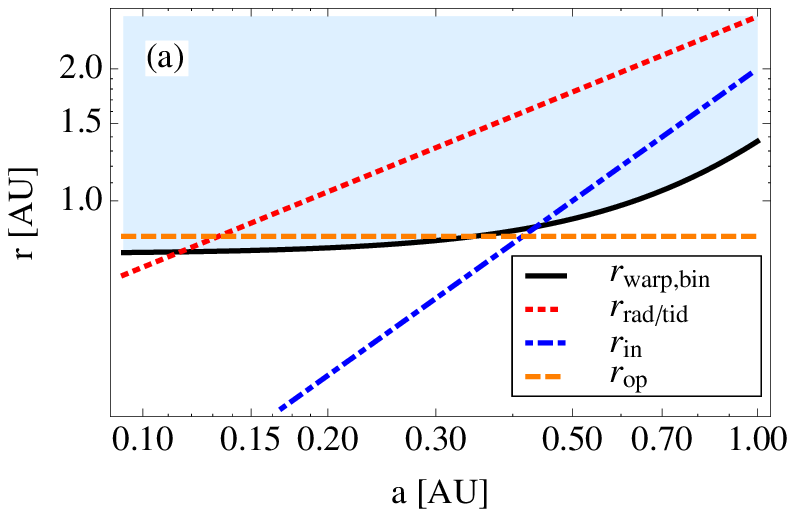}
\includegraphics[]{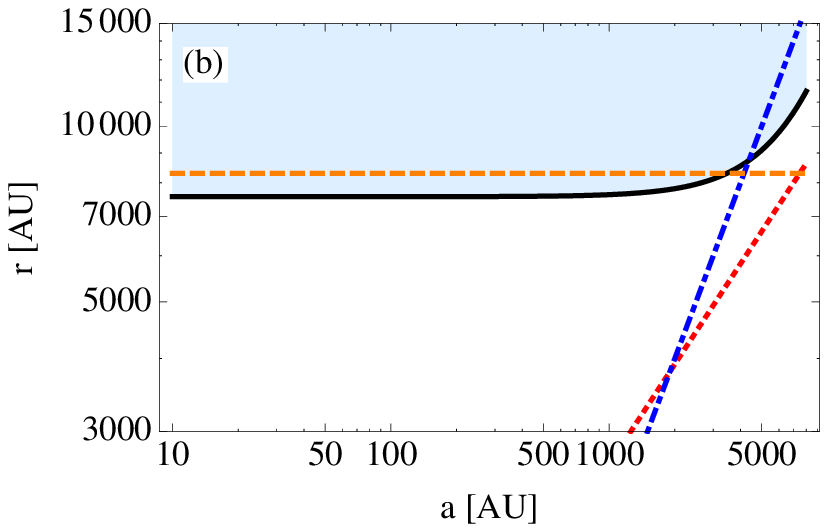}
}\\
\resizebox{\hsize}{!}
{
\includegraphics[]{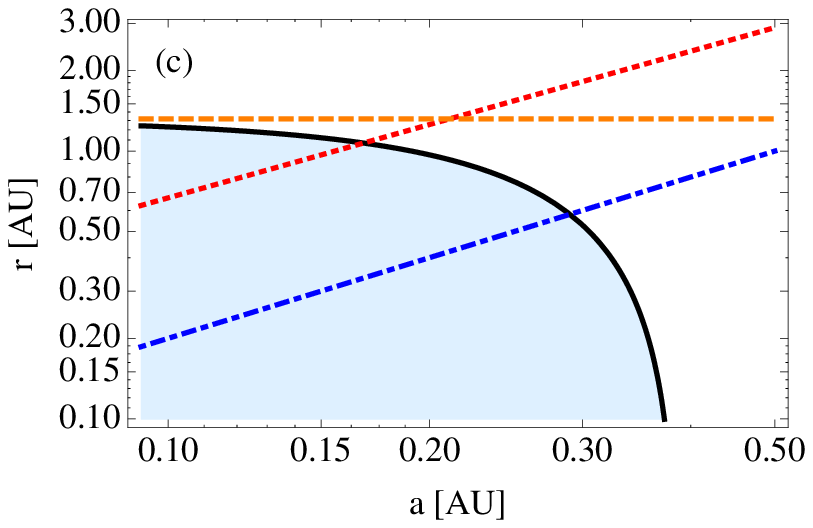}
\includegraphics[]{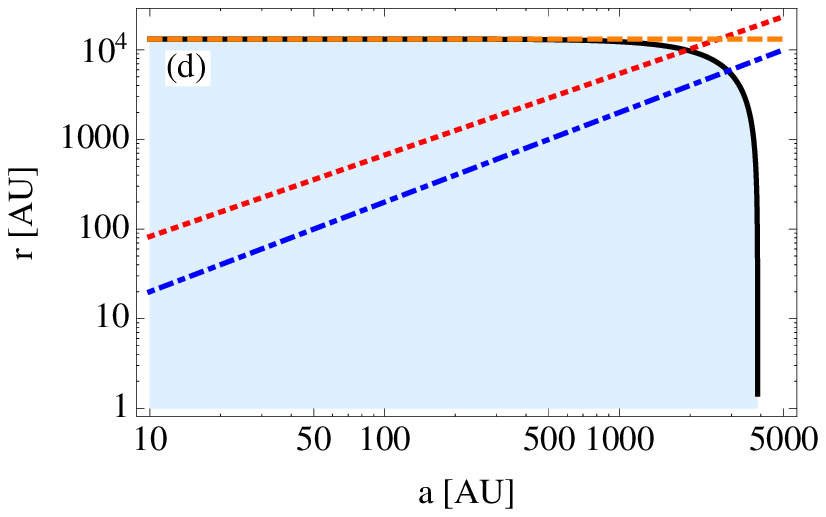}
}
\end{center}
\caption{
Characteristic radii of the warped circumbinary disk around young massive binaries on an eccentric orbit 
for $\alpha=0.01$, $s=-3/4$, $e=0.6$, $\zeta=1/4\,(q=1)$, $\xi_1\xi_2=1/4\,(q=1)$, $\gamma=\pi/4$, $T_0=100\,\rm{K}$, $\Sigma_{\rm{min}}=1\,{\rm{g\,cm^{-2}}}$, $M=30\,{M}_\odot$, and $L=10^4\,L_\odot$. While $r_0=1\,\rm{AU}$ is adopted in panels~(a) and (c), $r_0=0.1\,\rm{AU}$ is adopted in panels~(b) and (d). In panels~(a) and (b) $f=0.5$ and $n=-3/2$ are used, whereas $f=1$ and $n=-1/5$ are used in panels~(c) and (d).
In these four panels, the black solid line shows the marginally stable warping radius of the binary star case, which is given by equation~(\ref{eq:rwb}). The orange dashed line shows the outer radius of the optically thick region, which is given by equation~(\ref{eq:rop}). The red dotted line shows the tidal alignment radius where the growth timescale of the radiation-driven warping of the circumbinary disk is equal to the timescale during which the disk 
is aligned with the orbital plane by the tidal torques. The tidal alignment radius is given by equation~(\ref{eq:rradtid}). The blue dot-dashed line represents the inner radius of the circumbinary disk $r_{\rm{in}}=2\,a$.
%The shaded area between the blue solid and dashed lines represents the whole region of the circumbinary disk.
}
\label{fig:cradii}
\end{figure}
%%%%%%
%

%
% 6th paragraph
%
On the other hand, it is likely that a circumbinary disk around binary stars with much larger luminosities 
is unstable to radiation-driven warping.
Figure~\ref{fig:cradii} shows the dependence of the marginally stable warping radius in such a disk 
on the semi-major axis. Here, we adopt $\alpha=0.01$, $s=-3/4$, $e=0.6$, $\zeta=1/4\,(q=1)$, 
$\xi_1\xi_2=1/4\,(q=1)$, $\gamma=\pi/4$, $T_0=100\,\rm{K}$, $\Sigma_{\rm{min}}=1\,{\rm{g\,cm^{-2}}}$, 
$M=30\,{M}_\odot$, and $L=10^4\,L_\odot$.
While panels (a) and (c) are for $r_0=1\,\rm{AU}$, panels (b) and (d) are for $r_0=0.1\,\rm{AU}$. 
%In order to examine the effect of $f$, 
We also adopt $f=0.5$ and $n=-3/2$ in panels (a) and (b), and $f=1$ and $n=-1/5$ in panels (c) and (d).
%While $r_0=1\,\rm{AU}$ and $n=-3/2$ are adopted in panels~(a) and (c), $r_0=0.1\,\rm{AU}$ and $n=-1/5$ are adopted in panels~(b) and (d). While $f=0.5$ is adopted in panels~(a) and (b) , $f=1$ is adopted in panels~(c) and (d). 
In these four panels, the black solid line and orange dashed line show $r_{\rm{warp,bin}}$ and $r_{\rm{op}}$ in units of ${\rm{AU}}$, respectively. The red dotted line shows the radius where the growth timescale of the radiation-driven warping mode, $\tau_{\rm{rad}}$, equals the timescale for the disk to align with the orbital plane by the tidal torque, $\tau_{\rm{tid}}$. 
This tidal alignment radius is given by
\begin{eqnarray}
r_{\rm{rad/tid}}
=
%3.8\times10^2\,[{\rm AU}]
\left\{
9\pi{GM\Sigma_0}
\left(\frac{c}{L}\right)
\xi_1\xi_2
\sqrt{
5e^2(3e^2+2)\cos^2\gamma+(e^2-1)^2
}
\right\}^{1/(2-n)}
\left(
\frac{a}{r_0}
\right)^{2/(2-n)}
r_0,
%\nonumber \\
\label{eq:rradtid}
\end{eqnarray}
from equations~(\ref{eq:ttid}), (\ref{eq:trad}) and (\ref{eq:sigma}).
The growth of a finite-amplitude warping mode induced by the radiative torque can be significantly suppressed by the tidal torque in the region inside the tidal alignment radius. The blue dot-dashed line show the inner radius of the circumbinary disk. Here, the inner radius is assumed to be equal to the tidal/resonant truncation radius, where the tidal/resonant torque is balanced with the horizontal viscous torque of the circumbinary disk \citep{al94}. In the case of eccentric binaries with a nearly equal mass ratio, the tidal/resonant truncation radius is estimated to be $\sim2\,a$. The shaded area shows the region where the circumbinary disk is unstable to the warping modes induced by the radiative torques.

%
% 7th paragraph
%
From the figure, the marginally stable warping radius is on a sub-AU to kilo-AU scale for appropriate model parameters. In panes~(a) and (b), the circumbinary disk is warped in the region outside all of the marginally stable warping radius, the tidally alignment radius, and the inner disk radius, and inside the outer radius of the optically thick part (in the region above the black line, below the orange dashed line, and on the left-side of both the red dotted line and the blue dot-dashed line). On the other hand, in panels~(c) and (d), the circumbinary disk is warped in the region inside both the marginally stable warping radius and the outer radius of the optically thick region, and outside both the tidal alignment radius and the inner disk radius (in the region below the black line and the orange dashed line, and on the left-side of both the red dotted line and the blue dot-dashed line). Note that in any case the middle part of the circumbinary disk is warped because the warping mode in the inner part is suppressed by the tidal torques, and the radiative torques are effective only within the outer radius of the optically thick region. 

%
% 8th paragraph
%
The local precession frequency $\Omega_{\rm{p,tot}}$ of the linear warping mode is obtained from the dispersion relation by
\begin{eqnarray}
\Omega_{\rm{p,tot}}=\Omega_{\rm{p,tid}}+\Omega_{\rm{p.rad}}+\Omega_{\rm{p,vis}},
\label{eq:pfreq0}
\end{eqnarray}
where 
\begin{eqnarray}
\Omega_{\rm{p,tid}}&=&-\frac{3}{4}
\xi_1\xi_2
\sqrt{1+3e^2}
\left(\frac{a}{r}\right)^{7/2}
\Omega_{\rm{orb}},
\\
\Omega_{\rm{p,rad}}&=&\frac{3}{2}\zeta
\sqrt{1+3e^2}
\left(\frac{a}{r}\right)^{2}
\frac{1}{\tau_{\rm{rad}}},
\\
\Omega_{\rm{p,vis}}
&=&\frac{k\nu_2}{2r}(n+s+3),
%\frac{(n+s+3)}{\tau_{\rm{rad}}}\left[1-3\zeta\left(1+\frac{3}{2}e^2\right)\left(\frac{a}{r}\right)^2\right]^{1/2}
%\nonumber \\
%\lesssim\frac{(n+s+3)}{\tau_{\rm{rad}}}\left[1-\frac{3}{2}\zeta\sqrt{1+3e^2}\left(\frac{a}{r}\right)^2\right],
%\nonumber \\
\end{eqnarray}
where $\Omega_{\rm{p,tid}}$, $\Omega_{\rm{p,vis}}$, $\Omega_{\rm{p,rad}}$ are the local precession frequencies
due to the tidal, viscous, and radiative torques, respectively. Note that $\Omega_{\rm{p,tid}}$ has the same expression with the opposite sign to the precession frequency of the tidally induced $m=1$ wave (see equation~(11) of \citealt{khato09}). This satisfies the condition for simultaneous resonant excitation of eccentric and tilt waves proposed by \cite{kato14}. Therefore, both waves are likely to coexist in the circumbinary disk. 

\begin{figure}[ht!]
\resizebox{\hsize}{!}{
\includegraphics[width=14cm]{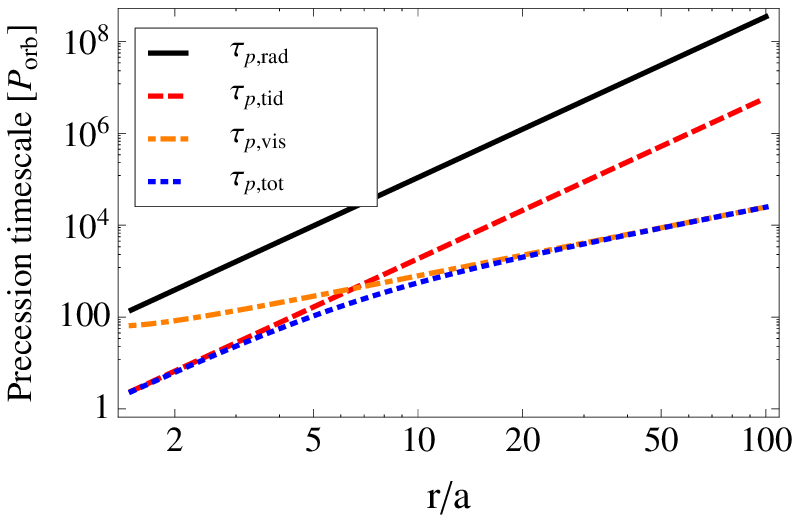}
}
\caption{Precession timescale of the warped circumbinary disk around young massive binaries
on an eccentric orbit with $e=0.6$, $\zeta=1/4\,(q=1)$, $\xi_1\xi_2=1/4\,(q=1)$, $\Sigma=1\,\rm{g\,cm^{-2}}$, $n=-3/2$, $s=-3/4$, $M=30\,{M}_\odot$, and $a=1\,{\rm{AU}}$. Each precession timescale is normalized 
by $P_{\rm{orb}}\simeq67\,{\rm{days}}$. The black solid line, red dashed line, orange dot-dashed 
line, and blue dotted line show the precession timescales from the radiative torque, from the tidal 
torque, from the viscous torque due to vertical shear, and from the sum of those three torques, 
respectively.
}
\label{fig:precess}
\end{figure}

%
% 9th (last) paragraph
%
Adopting $k\le(2\Gamma/\nu_2)\sqrt{1-3\zeta(a/r)^2(1+(3/2)e^2)}$ from equation~(\ref{eq:waven}), we obtain
\begin{eqnarray}
\Omega_{\rm{p,tot}}
&\lesssim&
-\frac{3}{2}
\sqrt{1+3e^2}
\left(\frac{a}{r}\right)^2\left[
\frac{1}{2}
\xi_1\xi_2
\Omega
+
\left(n+s+2\right)
\frac{\zeta}{\tau_{\rm{rad}}}
\right]
+\frac{n+s+3}{\tau_{\rm{rad}}},
\label{eq:pfreq}
\end{eqnarray}
%\begin{eqnarray}
%\Omega_{\rm{p,tot}}\approx\frac{3}{4}\xi_1\xi_2\left(1+\frac{3}{2}e^2\right)\left(\frac{a}{r}\right)^{7/2}\Omega_{\rm{orb}}.
%\end{eqnarray}
While the circumbinary disk tends to precess in the retrograde direction by the tidal torques, 
it tends to precess in the prograde direction by the other two torques. Figure~\ref{fig:precess} 
shows the radial dependence of the precession timescales.
The black solid line, the red dashed line, the orange dot-dashed line, and the blue dotted line show the precession timescale for the radiative torques  $\tau_{\rm{p,rad}}=1/\Omega_{\rm{p,rad}}$, tidal torques $\tau_{\rm{p,tid}}=1/\Omega_{\rm{p,tid}}$, viscous torque due to vertical shear $\tau_{\rm{p,vis}}=1/\Omega_{\rm{p,vis}}$, and total torque $\tau_{\rm{p,tot}}=1/\Omega_{\rm{p,tot}}$, respectively. Each precession timescale is normalized by the orbital period. The precession timescale is much longer than the orbital period for $r/a\gg1$. The circumbinary disk slowly precesses in the retrograde direction in the range of $r_{\rm{in}}/a\le{r}/a\le100$, because $\tau_{\rm{p,tid}}$ is shorter than the other two timescales.
%
%%%%%%%
% Figure 4
%%%%%%%
%

%
%%%%%%%%%%%%%%%%%
\section{Summary and Discussion}
\label{sec:4}
%%%%%%%%%%%%%%%%%
We have investigated the instability of a warping mode in a geometrically thin, non-self-gravitating 
circumbinary disk induced by radiative torques originated from two young stars in a binary on an 
eccentric orbit. For simplicity, they have been regarded as point irradiation sources. We have 
formulated the external torques acting on such a circumbinary disk, which are composed of both 
the tidal torques due to a time-dependent binary potential and the radiative torques. Based on these 
formulations, we have derived the warping mode induced by the radiative torques and compared 
the timescales of precession caused by tidal, viscous, and radiative torques for a small tilt angle. 
We have found that there is a marginally stable warping radius within or beyond which the circumbinary 
disk is unstable to radiation-driven warping, depending on the disk density and temperature gradient 
indices. Our main conclusions other than this instability condition are summarized as follows:
\begin{enumerate}
\renewcommand{\theenumi}{\arabic{enumi}}
\item The marginally stable warping radius is sensitive to the viscosity parameter ($\alpha$), the surface density measured at the radius where the disk changes from the optically thick to thin ($\Sigma_{\rm{min}}$), a fiducial radius ($r_0$),  the disk temperature at $r_0$ ($T_0$), and the stellar luminosity ($L$) [equation (44)], whereas it weakly depends on the orbital eccentricity and binary irradiation parameter, which is a function of binary mass ratio and luminosity of each star [equation (43)]. The marginally stable radius is on a sub-AU to kilo-AU scale for $\alpha=0.01$, $\Sigma_{\rm{min}}=1\,\rm{g\,cm^{-2}}$, $T_0=100\,\rm{K}$, $M=30\,M_\odot$ and $10^4\,L_\odot$, while it is much smaller than the stellar radius for $M=1\,M_\odot$ and $L=1\,L_\odot$. The radiation-driven warping can, therefore, be a plausible mechanism for a warped structure of a circumbinary disk around a young massive binary, whereas it is an unlikely mechanism for that around a young low-mass binary.
\item There is a clear difference in the warping radius between the single star case and the binary star case. 
Since the tidal torques work on the circumbinary disk so as to align the disk plane with the binary orbital plane, they can suppress finite-amplitude warping modes induced by the radiative torques. The circumbinary disk, therefore, starts to be warped at the tidal alignment radius where the growth timescale of the radiation-driven warping is equal to the timescale of the disk alignment due to tidal torques, if the tidal alignment radius is 
larger than the marginally stable warping radius. In contrast, the circumstellar disk around a single star starts to be warped in the region beyond the marginally stable warping radius.
%between the inner disk radius and the marginally stable warping radius.
%
\item The circumbinary disk precesses by tidal torque, radiative, and vertical viscous torques. 
While the radiative and vertical viscous torques tend to precess the circumbinary disk in the 
prograde direction, the tidal torques tend to precess it in the retrograde direction. Since the 
former two precession frequencies are substantially lower than the latter precession frequency, 
the circumbinary disk slowly precesses in the retrograde direction. The precession timescale 
is much longer than the orbital period for the outer part of the circumbinary disk ($r/a\gtrsim4$). 
Therefore, it is unlikely that the periodic light variation due to the precession of the warped disk 
could be detected.
\item The tidal torques with a small tilt angle depend on the azimuth of the tilt for $e\neq0$, 
whereas they have no such dependence for $e=0$.
%
%\item The binary irradiation parameter depends only on the binary mass ratio, if the stellar mass-luminosity relationship is adopted, and then is up to $1/4$ for an arbitrary binary mass ratio. This is different from the case studied in Paper I where the irradiation sources are two accretion disks.
%
\item The higher the orbital eccentricity, the longer the growth timescale of radiation-driven warping mode and the shorter its precession timescale. The effect is, however, significant only in the region $r/a<2$. For $r/a\gtrsim{3}$, the growth timescale of the warping mode in the binary star case is reduced to that of the single star case.
\end{enumerate}

%
% 2nd paragraph
%
For simplicity, we have assumed that the circumbinary disk is initially aligned with 
the binary orbital plane ($\beta\ll1$), as in most of the previous studies. However, the 
angular momentum vector of the circumbinary disk does not always coincide with that 
of the binary orbital angular momentum, because the orientation of the circumbinary disk 
is primarily due to the angular momentum of material accreted after the binary has formed. 
Therefore, the orientation of the circumbinary disk plane can be taken arbitrarily with respect 
to the binary orbital plane. In a misaligned system with a significant tilt angle, the inner part 
of the circumbinary disk tends to align with the binary orbital plane by the tidal interaction 
between the binary and the circumbinary disk, whereas the outer part tends to retain the 
original state by the shear viscosity in the vertical direction. 
As a result, the circumbinary disk should be warped without the effect of radiation driven 
warping instability (e.g., \citealt{flp13, lf13,fl14}).
It is important to examine how the radiation driven warping instability 
works in the misaligned systems under the tidal potential, but it is difficult to find the analytic 
solutions because of the complicated dependence of  the tidal and radiative torques on the 
tilt angle and azimuth of tilt. We will numerically study this problem in the future.

%
% 3rd paragraph
%
%There is a cavity between the circumbinary disk and the binary (see Figure~\ref{fig:schmatic}), which is elongated even in a circular binary case because of the binary-disk interaction (e.g., \citealt{mm08}). The inner radius of the circumbinary disk, i.e., the outer radius of the cavity, is equal to the tidal truncation radius, where the tidal torque is balanced with the viscous torque of the circumbinary disk, and is typically $\sim 2a$. Since the marginally stable warping radius is substantially larger than the inner radius of the circumbinary disk, the shape of the cavity gives little influence on the warping condition.

%
% 3rd (last) paragraph
%
There is a cavity between the circumbinary disk and the binary for a nearly equal mass to moderate 
mass ratio (see Figure~\ref{fig:schmatic}) \citep{al94}. In the case of an extreme mass ratio ($q\ll1$), 
however, the material accretes through the cavity and therefore the circumbinary disk cannot maintain 
its cavity any longer because of significantly weaker tidal torque than in a moderate mass ratio binary. 
In such a situation, the system is composed of two circumstellar disks around individual young stars: a 
substantially larger accretion disk around the more massive, primary star (the primary disk) and a smaller 
accretion disk around the less massive, secondary star (the secondary disk), although there are some 
observational cases, in which the secondary disk is more massive than the primary disk \citep{aj14}.
In the system, the tidal torque of the secondary star tends to align the outer part of the primary disk with 
the orbital plane. Since the primary disk is unstable to radiation-driven warping beyond/within the marginally 
stable warping radius, only the outer/inner part of the primary disk is warped. Note that the marginally 
stable warping radius is smaller than the stellar radius for binaries with young low-mass stars with 
$\sim1\,M_\odot$ and $\sim1\,L_\odot$ from equation~(\ref{eq:rws}), as far as $\Sigma_{\rm min}\ge10^{-3}\,{\rm g\,cm^{-2}}$, 
or, the disk opacity to the irradiation from the two stars less than $10^{3}\,\rm{cm^{2}\,g^{-1}}$. Therefore, 
the discussion on the possible radiation-drivien warping will be applied only to massive star cases. 
In addition to this, the primary disk is truncated by the tidal torque due to the secondary star. There are 
thus three characteristic radii in a circumstellar disk: the marginally stable warping radius, the tidal alignment 
radius, and the tidal truncation radius. The situation is clearly different from the circumbinary case. We will 
examine the warping instability in circumstellar disks in a subsequent paper.

%
% 4th (last) paragraph
%
According to the recent observations, the cirsumbinary and circumstellar disks in pre-main sequence 
binaries are likely to be composed of the inner gas disk and outer dusty disk. The dust opacity is much larger 
than the gas opacity in the case of the wavelength less than $100\,\mu{m}$ \citep{oh94}. If the dust opacity is more than 
$10^4\,{\rm cm^{2}g^{-1}}$, the warping radius is comparable or larger than $1\,{\rm AU}$ even if the stellar 
luminosity is as low as $\sim1\,L_\odot$. This suggests that a circumbinary or circumstellar disk in a low-mass 
young binary system could be warped by the radiative torque. Furthermore, such a dusty disk should be flared 
in the vertical direction \citep{cg97}. If this is the case, as the tilt angle locally increases, the radiative torques from the stars 
are stronger and hence the marginally stable warping radius is closer to the stars than in geometrically thin disk cases. 
We will also investigate in future the effect of radiation driven warping on circumbinary or circumstellar dusty disks.

%\appendix
%
%%%%%%%%%%%%%%
\section*{Acknowledgments}
%%%%%%%%%%%%%%
%
The authors thank to the anonymous referee for fruitful comments and suggestions.
K.H. is grateful to Jongsoo~Kim for helpful discussions and his continuous encouragement. 
K.H. would also like to thank the Kavli Institute for Theoretical Physics (KITP) for their hospitality 
and support during the program on "A Universe of Black Holes". B.W.S. and T.H.J. are grateful for 
support from KASI-Yonsei DRC program of Korea Research Council of Fundamental Science 
and Technology (DRC-12-2-KASI). This work was also supported in part by the Grants-in-Aid for 
Scientific Research (C) of Japan Society for the Promotion of Science (23540271 T.N. and K.H.; 
24540235 A.T.O. and K.H.).

%
%%%%%%%%%%%%%

\end{document}